\documentclass[twocolumn,showpacs, prl,nobalancelastpage,floatfix]{revtex4-1}
\usepackage{float}
\usepackage{epsfig}
\usepackage{amsmath}
\usepackage{graphicx}
\usepackage{dcolumn}
\usepackage{latexsym}
\usepackage{caption}
\usepackage{subcaption}

\usepackage{slashed}
\begin{document}
\title{\Large next-nearest-neighbor tight-binding model of plasmons in graphene}

\author{V. Kadirko$^1$, K. Ziegler$^2$ and   E. Kogan$^1$}

\address{$^1$ The Department of Physics, Bar--Ilan University, Ramat Gan 52900, Israel \\
$^2$ Institute fur Physik, Universitat Augsburg}

%
%
\begin{abstract}
In this paper we investigate the influence of the next-nearest-neighbor coupling of
tight-binding model of graphene on the spectrum of plasmon excitations. The nearest-neighbor
tight-binding model was previously used to calculate plasmon spectrum in the next paper \cite{ziegler} . We expand the previous results of the paper by the next-nearest-neighbor tight-binding model. Both methods are based on the numerical calculation of the dielectric function of graphene and loss function. Here we compare plasmon spectrum of the next-nearest and nearest-neighbor tight-binding models and find differences between  plasmon dispersion of two models.
\end{abstract}

\date{\today}
\maketitle

%
%
\section{1 introduction}

Graphene, a single layer of carbon atoms arranged as a honeycomb lattice, is a semimetal with remarkable physical
properties \cite{kastro,abergel}. This is due to the band structure of the material which consists of two bands
touching each other at two nodes. The electronic spectrum around these two nodes is linear and can be approximated by Dirac cones.
However, calculations of many physical properties demand the knowledge of the full electron dispersion in the entire Brillouin zone,
not only in the vicinity of the nodes. This statement becomes particularly relevant when we take into account
the fact that graphene can be gated or doped, such that the Fermi energy can be freely tuned.

One of the main open issues in the physics of graphene is the role played by electron-electron
interaction. In doped graphene long range Coulomb interaction leads to a gapless plasmon mode which can be described
theoretically within the random phase approximation (RPA). Although this is a standard problem in semiconductor physics, it was
studied initially in the case of graphene only in the Dirac approximation around the nodes \cite{wunsch,hwang,polini}.
The linear approximation leads to a frequency of the plasmon that is proportional to the square root of the wavevector.

Later the plasmon dispersion law was also calculated
for the more realistic band structure, obtained in the framework of the tight-binding model with nearest-neighbor hopping
\cite{ziegler,katsnelson}.
This model is characterized by two symmetric bands, which implies a chiral symmetry. The latter connects the eigenstates
of energy $-E$ directly with eigenstates of energy $E$ by a linear transformation. This symmetry, which also realizes a
particle-hole symmetry, is broken by a next-nearest-neighbor hopping term. Usually, physically properties change
qualitatively under symmetry breaking. Here we would like to study the effect of particle-hole symmetry breaking due to
next-nearest-neighbor hopping on the plasmon dispersion.
For this purpose we extend the nearest-neighbor hopping approximation used in \cite{ziegler} by taking into account
the next-nearest-neighbor hopping.

We consider an electron gas which is subject to an electromagnetic potential \( V_{i}(\textbf{q}, \omega)\). The response of the electron gas is to create
a screening potential \( V_{s}(\textbf{q}, \omega)\) which is caused by the rearrangement of the electrons due to the external potential.
Therefore, the total potential, acting on the electrons, is\cite{Mahan}
\begin{align}
V(\textbf{q}, \omega) =  V_{i}(\textbf{q}, \omega) +  V_{s}(\textbf{q}, \omega).
\end{align}
\(V_s\) can be evaluated self-consistently  \cite{Ehrenreich} and is expressed via the dielectric function \(\epsilon(\textbf{q},\omega)\). Then the total potential reads \cite{Mahan}
\begin{align}
V(\textbf{q}, \omega)  = \frac{1}{\epsilon(\textbf{q},\omega)}  V_{i}(\textbf{q}, \omega).
\end{align}

%
%
\section{2 Nearest and next-nearest hopping model}

The tight-binding Hamiltonian for electrons in
graphene with  both
nearest- and next-nearest-neighbor hopping has the form\cite{kastro}
(we use units such that \( \hbar = 1 \))
\begin{align}
H = &-t\sum_{<i,j>,\sigma} (a_{\sigma,i}^\dagger b_{\sigma,j} +H.c.)  \nonumber \\
  &-t'\sum_{<<i,j>>,\sigma} (a_{\sigma,i}^\dagger a_{\sigma,j} + b_{\sigma,i}^\dagger b_{\sigma,j} + H.c.)
\end{align}
where \( a_{\sigma,i}(a_{\sigma,i}^{\dagger}) \) annihilates (creates) an electron with spin \( \sigma (\sigma= \uparrow , \downarrow)\) on site
\( \textbf{R}_i\) on sublattice A (an equivalent definition is used for sublattice B), $t \approx 2.8$ eV
is the
nearest-neighbor hopping energy (hopping between different
sublattices), and \(t'\) is the next nearest-neighbor
hopping integral  (hopping in the same sublattice).
The value of $t'$ is not well known but ab initio calculations
find $0.02t \leq t'\leq 0.2t$ depending on the tight-binding
parametrization \cite{kastro}.

The matrix representation of the Hamiltonian is
\begin{align}
H =
 \begin{pmatrix}
 h_0  & h_1 - i h_2 \\ h_1 + i h_2 & h_0
\end{pmatrix}.
\label{ham0}
\end{align}
The non-diagonal terms in the Hamiltonian correspond to the nearest-neighbor hopping\cite{ziegler}:
\begin{equation}
\label{eq:h1h2}
h_1 = -t \sum_{j=1}^3 \cos(\textbf{b}_j \textbf{k}), \quad   h_2 = -t \sum_{j=1}^3 \sin(\textbf{b}_j \textbf{k}),
\end{equation}
where \( \textbf{b}_{1,2,3} \) are the nearest-neighbor vectors  on the honeycomb lattice:
$\textbf{b}_1 = d(1/2, \sqrt{3}/2),\textbf{b}_2 = d(1/2, -\sqrt{3}/2)  \textbf{b}_3 = d(-1, 0)$ and \( d \) is the lattice constant (\( \approx \) 1.42  \AA).
The diagonal terms correspond to next-nearest-neighbor hopping:
\begin{equation}
h_0 = -t' \sum_{j=1}^6 \cos(\textbf{a}_j \textbf{k}),
\end{equation}
where $\textbf{a}_1 = d(3/2, \sqrt{3}/2), \textbf{a}_2 = d(-3/2, -\sqrt{3}/2), \textbf{a}_3 = d(3/2, -\sqrt{3}/2),
\textbf{a}_4 = d(-3/2, \sqrt{3}/2), \textbf{a}_5 = d(0, -\sqrt{3}),\textbf{a}_6 = d(0, \sqrt{3})$.

\begin{figure}
\centering
\includegraphics[width=0.5\textwidth]{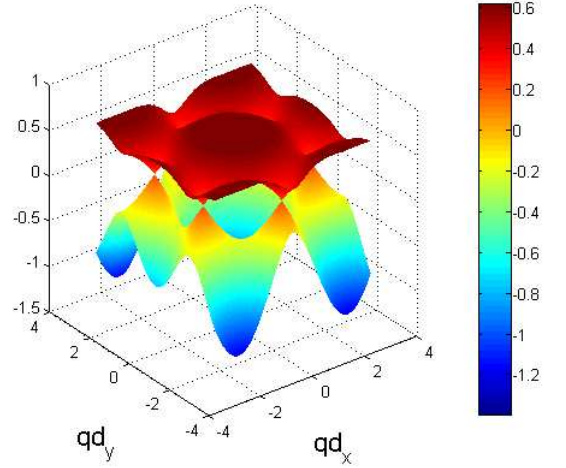}

\caption{Energy dispersion  of graphene with $t' = 0.2 t$. The figure shows a broken particle--hole symmetry. The Dirac nodes are shifted
 up by \(3 t'\). The energy is measured in the units of the electronic bandwith \(\Delta = 3 t\).  }
\label{fig:dispersion}
\end{figure}

The
energy bands derived from this Hamiltonian have the
form\cite{kastro}
\begin{align}
\label{eq:dispersion}
E_{\pm}({\bf k}) &= \pm t \sqrt{3 + f(\textbf{k})} - t' f(\textbf{k}), \nonumber \\
f(\textbf{k}) &= 2\cos(\sqrt{3}k_y a) + 4\cos(\frac{\sqrt{3}}{2}k_y a)\cos(\frac{3}{2}k_y a)
\end{align}
where the plus sign applies to the upper (\(\pi\) or conduction) and the minus sign the lower (\(\pi*\)
or valence) band.
It should be noticed that the presence of \(t'\) shifts the position of the Dirac point in energy and
it breaks electron-hole symmetry.
In both cases, nearest-neighbor and next nearest-neighbor hopping, the electronic dispersion is an even function\cite{ziegler}
\begin{align}
  E_{\textbf{k},l} = E_{-\textbf{k},l} \ .
\end{align}
The dispersion law for next nearest-neighbor hopping is presented on Fig. \ref{fig:dispersion},
 and the  eigenvectors of the Hamiltonian read
\begin{equation}
\frac{1}{\sqrt{2}}
 \begin{pmatrix}
 \frac{h_1 - i h_2}{\sqrt{h_1^2 + h_2^2}} \\ 1
\end{pmatrix},
\frac{1}{\sqrt{2}}
 \begin{pmatrix}
 \frac{h_1 - i h_2}{\sqrt{h_1^2 + h_2^2}} \\ -1
\end{pmatrix}
\end{equation}
where the first eigenvector is for the upper band and the second eigenvector for the lower band.

The Hamiltonian $H$ in Eq. (\ref{ham0}) has a chiral symmetry for $h_0=0$:
\begin{equation}
e^{\alpha\sigma_3}He^{\alpha\sigma_3}=H , \ \ \ \sigma_3=
\begin{pmatrix}
1 & 0 \\ 0 & -1
\end{pmatrix}
\ ,
\label{chiral0}
\end{equation}
which connects eigenstates of energy $-E$ with eigenstates of energy $E$ by
\begin{equation}
\Psi_{-E}=\sigma_3\Psi_E
\ .
\end{equation}
This is not the case after we have broken the chiral symmetry by the next-nearest-neighbor hopping
term $h_0$.

\section{3 Dielectric Function}

The longitudinal dielectric function in calculated in RPA\cite{Dressel,Ehrenreich}:
 \begin{eqnarray}
 \label{eq:dielectric}
&&\epsilon(\textbf{q},\omega) = 1 - \frac{2\pi e^2}{q \kappa}\chi(\textbf{q},\omega)
\end{eqnarray}
where \(\kappa\) is a dielectric constant and \(\chi\) is a polarizability.
For polarizability we used the Lindhard formula  \cite{Dressel}, which in our case
after some straightforward calculations can be reduced to the expression
\begin{equation}
\chi(\textbf{q}, \omega) =\chi_1(\textbf{q}, \omega) +\chi_2(\textbf{q}, \omega)
\end{equation}
with the intraband contribution
\begin{eqnarray}
\label{eq:zpolarizability}
&&\chi_1(\textbf{q}, \omega) = \lim_{\delta \to 0} \sum_{s,t = \pm 1}
\int\limits_{BZ} \frac{1}{4}|\kappa_{\textbf{k}}^* \kappa_{\textbf{k+q}} + 1 |^2 \\
&& \times \frac{f(sE_{\textbf{k}}^{(1)}+E_{\textbf{k}}^{(2)})}{s\left(E_{\textbf{k}}^{(1)}
- E_{\textbf{k + q}}^{(1)}\right)+ E_{\textbf{k}}^{(2)}-E_{\textbf{k + q}}^{(2)} + t(\omega - i \delta)} d^2 k,
\nonumber
\end{eqnarray}
and the interband contribution
\begin{eqnarray}
\label{eq:zpolarizability2}
&&\chi_2(\textbf{q}, \omega) = \lim_{\delta \to 0} \sum_{s,t = \pm 1}
\int\limits_{BZ} \frac{1}{4}|\kappa_{\textbf{k}}^* \kappa_{\textbf{k+q}} - 1 |^2 \\
&& \times \frac{f(sE_{\textbf{k}}^{(1)}+E_{\textbf{k}}^{(2)})}{s\left(E_{\textbf{k}}^{(1)}
+ E_{\textbf{k + q}}^{(1)}\right)+ E_{\textbf{k}}^{(2)}-E_{\textbf{k + q}}^{(2)}+ t(\omega - i \delta)} d^2 k,
\nonumber
\end{eqnarray}
where
\begin{eqnarray}
&&\kappa_{{\bf k}} = (h_1 - i h_2) / E_{{\bf k}},\\
&&h_1 = -t\left[\cos(k_xd)\cos(\sqrt{3}k_yd)+\cos(k_sd)\right],\nonumber\\
&&h_2 = -t\left[\sin(k_xd)\cos(\sqrt{3}k_yd)+\sin(k_sd)\right];\nonumber
\end{eqnarray}
\( f(E) = 1/(e^{\beta (E - \mu)} + 1)\) is the Fermi-Dirac distribution function,
 \( \beta = 1/k_B T\),  \(\mu\) is a  chemical potential.
The energies are defined as
\begin{eqnarray}
&&E_{\textbf{k},l} = E_{\textbf{k},l}^{(1)} + E_{\textbf{k},l}^{(2)}, \\  \nonumber
&&E_{\textbf{k},l}^{(1)} = (-1)^l \sqrt{h_1^2 + h_2^2} = \pm t \sqrt{3 + f(\textbf{k})}, \\ \nonumber
&&E_{\textbf{k},l}^{(2)} = -t' f(\textbf{k})
\end{eqnarray}
If we take \( E_{\textbf{k}}^{(2)} = E_{\textbf{k+q}}^{(2)} = 0 \) the integral yields the same polarizability formula as that found in the
nearest-neighbor model's polarizability \cite{ziegler,katsnelson}.

\section{4 plasmons in graphene}
In a first approximation, we can consider plasmons as collective excitations of electrons,
where the dielectric function vanishes\cite{Mahan}:
\begin{eqnarray}
\label{eq:diel_cond}
\epsilon (\textbf{q},\omega)= 0.
\end{eqnarray}
In general, however, the dielectric function is complex due to poles in the integrals
(\ref{eq:zpolarizability}), (\ref{eq:zpolarizability2}). This implies that (\ref{eq:diel_cond})
has no solution, unless we only request that the real part of the dielectric function vanishes:
\begin{eqnarray}
\label{eq:r_diel}
Re[\epsilon (\textbf{q},\omega)]= 0,
\end{eqnarray}
assuming a real function $\omega({\bf q})$ as the plasmon dispersion.
For a numerical evaluation of the integrals it is more convenient to consider the loss function\cite{polini,Mahan,Dressel}
\begin{eqnarray}
\label{eq:loss_f}
&&Im \Bigg ( \frac{1}{\epsilon(\textbf{q},\omega)} \Bigg ) = \frac{- Im [\epsilon(\textbf{q},\omega)]}{\{Re [\epsilon(\textbf{q},\omega)]\}^2
+ \{Im[\epsilon(\textbf{q},\omega)]\}^2}\nonumber\\
\end{eqnarray}
whose broadened peak indicates the plasmon. Here a complex solution $\omega({\bf q})$ gives both the dispersion from the real part
and the decay of the plasmons from the imaginary part.

\begin{figure}
\centering
\includegraphics[width=0.4\textwidth]{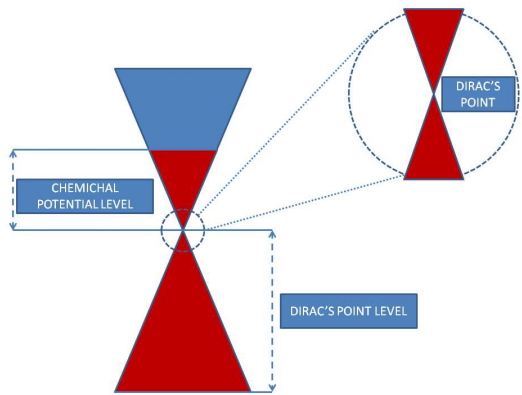}

\caption{The scheme shows that chemical potential $\mu$ is taken relative to Dirac cone. Dirac point's level is relative to case
  $t' = 0$ and the level equals to $ 3 t'$ .}
  \label{fig:DiracLevel}
\end{figure}

In the present paper
the polarizability of graphene \( \chi \) is evaluated numerically and the corresponding dielectric function is obtained from
Eq. (\ref{eq:dielectric}) for different values of the real frequency \(\omega \), the wave vector ${\bf q}$ and chemical potential (Fermi energy)
$\mu$. Moreover, we assume $\kappa=4$.
The chemical potential  level $\mu$ is selected to be relative to Dirac points whose existence is not affected
by a variation of the parameter $t'$ but are shifted by $ 3 t'$  (ref. \cite{wallace}), as shown in Fig. [~\ref{fig:DiracLevel}].\\

\begin{figure}[t!]
\centering
    \begin{subfigure}[b]{0.5\textwidth}
        \centering
        \includegraphics[width=\textwidth]{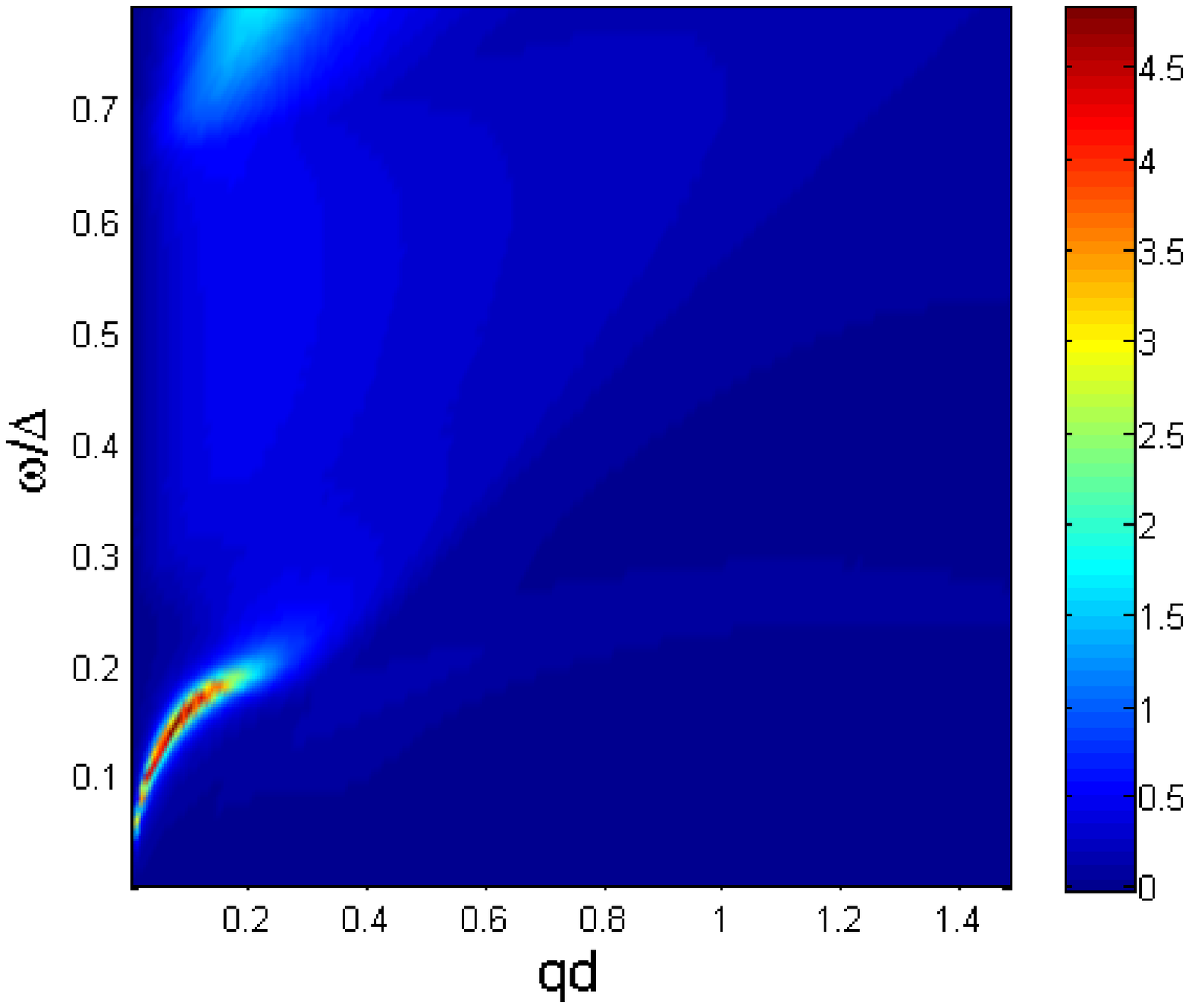}
        \caption{Plasmon dispersion for $\mu = 0.4 t, t' = 0.0 t$ and different values of  $q_y$ component ($q_x=0$)}
        \label{fig:comp1}
    \end{subfigure}


    \begin{subfigure}[b]{0.5\textwidth}
        \centering
        \includegraphics[width=\textwidth]{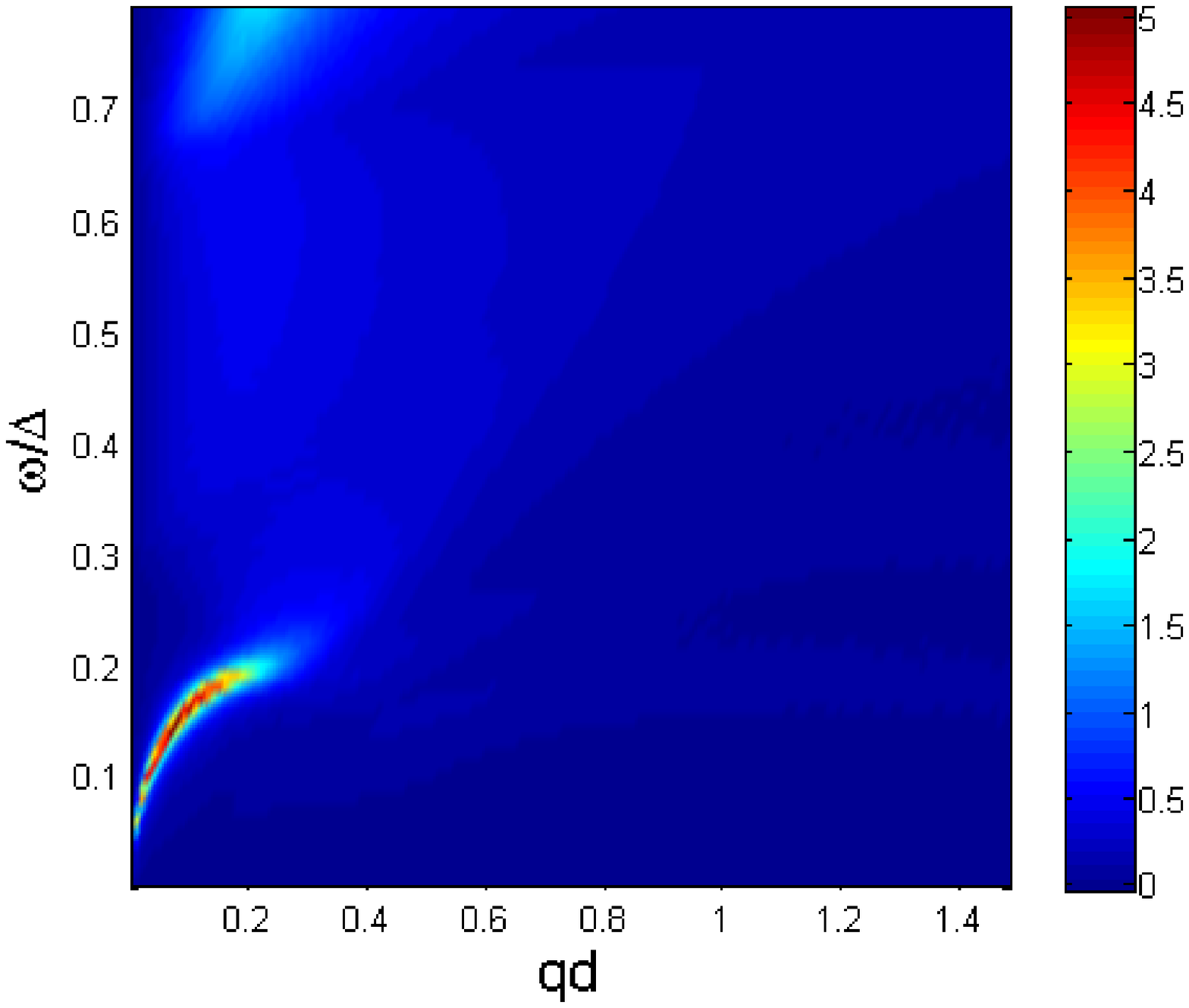}
        \caption{Plasmon dispersion for $\mu = 0.4 t, t' = 0.2 t$ and different values of  $q_y$ component ($q_x=0$)}
        \label{fig:comp1}
    \end{subfigure}

  \caption{Plasmon dispersion for $\mu = 0.4 t$}
  \label{fig:result1}
\end{figure}

\begin{figure}[h!]
\centering
    \begin{subfigure}[b]{0.5\textwidth}
        \centering
        \includegraphics[width=\textwidth]{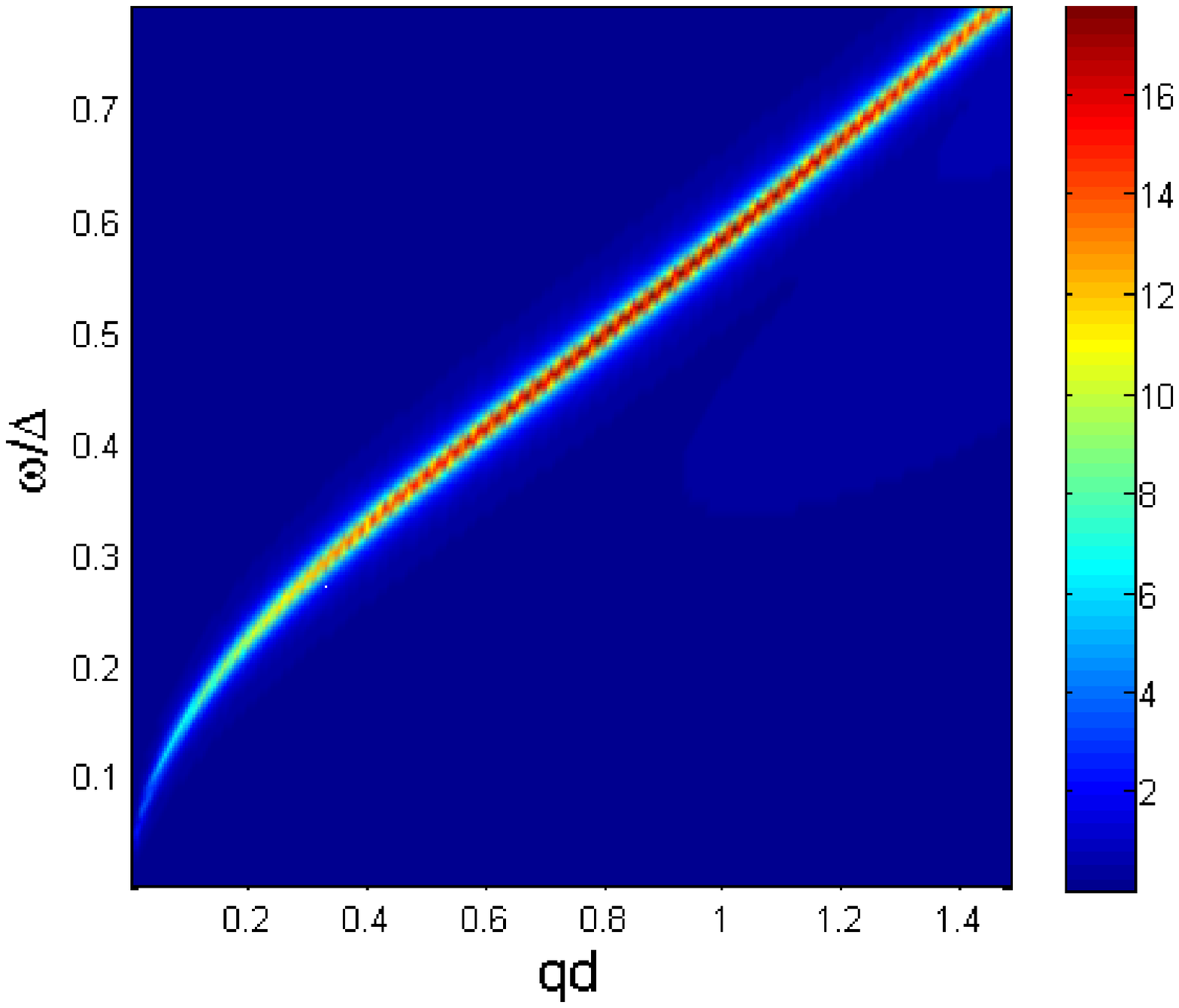}
        \caption{Plasmon dispersion for $\mu = -2.7 t, t' = 0.0 t$ and different values of  $q_y$ component ($q_x=0$)}
        \label{fig:comp1}
    \end{subfigure}


    \begin{subfigure}[b]{0.5\textwidth}
        \centering
        \includegraphics[width=\textwidth]{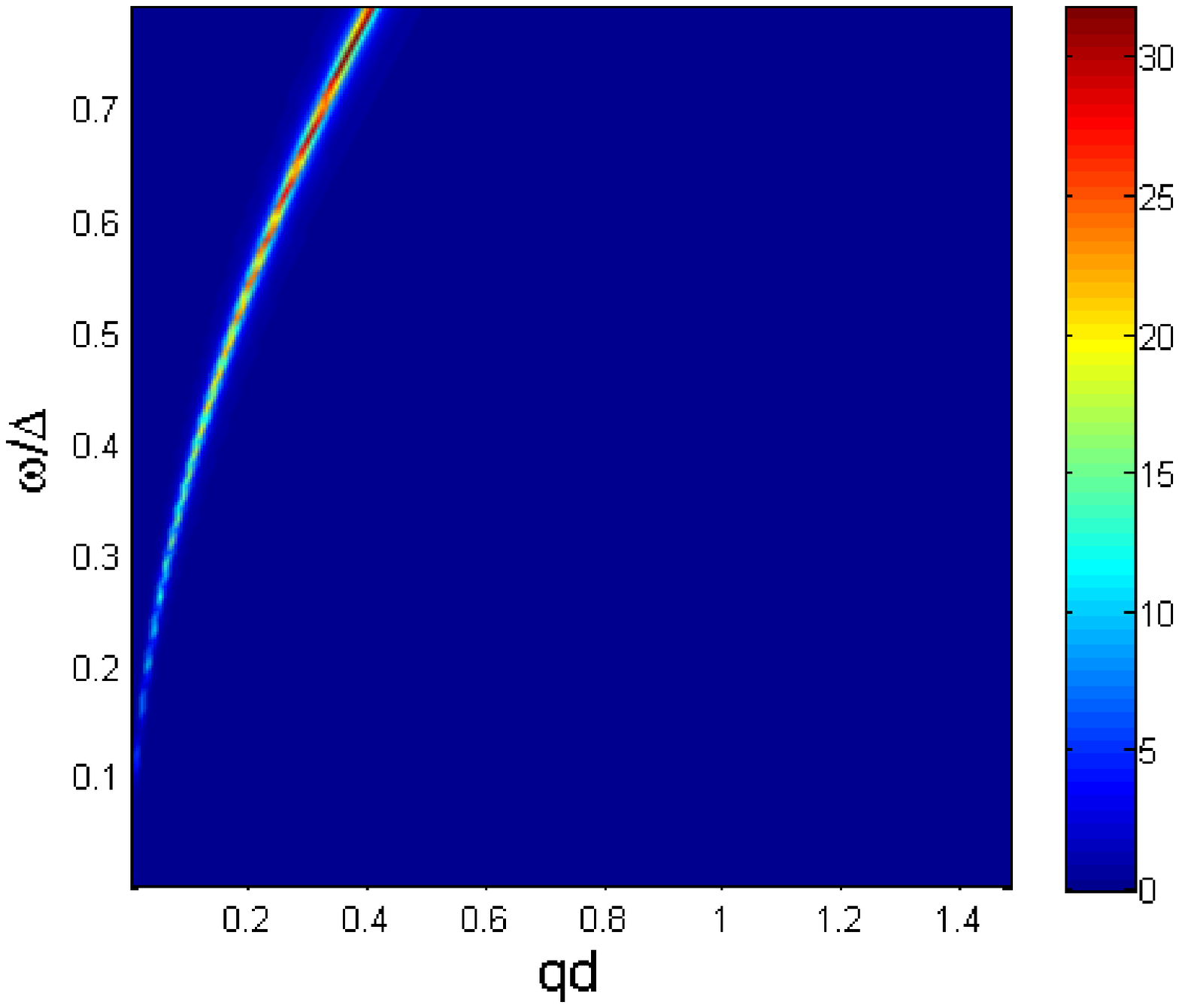}
        \caption{Plasmon dispersion for $\mu = -2.7 t, t' = 0.2 t$ and different values of  $q_y$ component ($q_x=0$)}
        \label{fig:comp1}
    \end{subfigure}

  \caption{Plasmon dispersion for $\mu = -2.7 t$}
  \label{fig:result3}
\end{figure}

Our results for plasmon dispersion law are shown in Figs. \ref{fig:result1} 
and \ref{fig:result3}.
For each figure we have selected two values for $t'$, namely $t' = 0 $ and $t' = 0.2 t $. The original chemical potentials
$\mu$ that appear in Figs. \ref{fig:result1} 
and \ref{fig:result3}
are taken from the previous paper \cite{ziegler} and are modified by the value $t' = 0.2 t$. 

The influence of next-nearest hopping parameter $t'$ is insignificant for the plasmon dispersion law when the chemical potential is above Dirac point,
as depicted in Figs. \ref{fig:result1} and \ref{fig:result3}. The shape of the plasmon dispersion law in Fig. \ref{fig:result1}
does not change significantly by a variation of the parameter $t' = 0, 0.2 t$. On the other hand, the result is quite different
when the chemical potential is below the Dirac point.
Fig. \ref{fig:result3} shows that for different values of hopping parameter $t'$ and for a negative chemical potential the shape of dispersion law
changes strongly and the dispersion curve is much sharper when the value of the hopping parameter is larger. In general, our calculations of the
plasmon dispersion law show that there is almost no
change of the plasmon dispersion with $t'$ when chemical potential is above the Dirac point.

%
%
\section{5 conclusion}

In conclusion we have investigated  the 2D tight-binding hamiltonian model
under the influence of the next nearest-neighbour coupling (constant) and
we theoretically obtained analytic expression for improved graphene polarizability
expression . Our work is extension to previous results obtained by \cite{ziegler}, where
only near-neighbor constant model is used. This work improves the previous results
for graphene plasmon's dispersion law.

The research of next-nearest hopping tight-binding model gave the possibility to investigate the plasmon's dispersion
law near Dirac point in the case of low values of chemical potential relative to Dirac point, by using
analytical calculations and numerically to show that dispersion's laws in two cases (near neighbor and next-nearest tight binding model) are almost the same as predicted theoretically.

%
%

%
%

\end{document}